\documentclass[journal]{IEEEtran}
\usepackage{graphicx}
\usepackage{amsmath}
\usepackage{amssymb}
\usepackage{booktabs}
\usepackage{algorithm}
\usepackage{algpseudocode}
\usepackage{tikz}
\usetikzlibrary{shapes.geometric, arrows.meta, positioning, fit, backgrounds, calc, shadows}
\usepackage{hyperref}
\usepackage{xcolor}
\usepackage{subcaption}
\usepackage{multirow}
\usepackage{url}

\begin{document}

\title{Grid-Mind: An LLM-Orchestrated Multi-Fidelity Agent\\for Automated Connection Impact Assessment}
\author{Mohamed Shamseldein,~\IEEEmembership{Senior Member,~IEEE}
\thanks{M. Shamseldein is with the Department of Electrical Power and Machines,
Faculty of Engineering, Ain Shams University, Cairo 11517, Egypt
(e-mail: mohamed.shamseldein@eng.asu.edu.eg).}
\thanks{Manuscript received February 2026. A provisional patent application (U.S. App. No. 63/989,282) covering the architectures and methods described herein is pending.}}

\IEEEoverridecommandlockouts
\IEEEpubid{\makebox[\columnwidth]{\parbox{\columnwidth}{\centering \textbf{PREPRINT: THIS WORK IS UNDER REVIEW\\FOR PUBLICATION IN IEEE TRANSACTIONS ON SMART GRID.}}}\hspace{\columnsep}\makebox[\columnwidth]{ }}

\maketitle

\begin{abstract}
Large language models (LLMs) have demonstrated remarkable tool-use capabilities, yet their application to power system operations remains largely unexplored. This paper presents Grid-Mind, a domain-specific LLM agent that interprets natural-language interconnection requests and autonomously orchestrates multi-fidelity power system simulations. The LLM-first architecture positions the language model as the central decision-making entity, employing an eleven-tool registry to execute Connection Impact Assessment (CIA) studies spanning steady-state power flow, N-1 contingency analysis, transient stability, and electromagnetic transient screening. A violation inspector grounds every decision in quantitative simulation outputs, while a three-layer anti-hallucination defense mitigates numerical fabrication risk through forced capacity-tool routing and post-response grounding validation. A prompt-level self-correction mechanism extracts distilled lessons from agent failures, yielding progressive accuracy improvements without model retraining.

End-to-end evaluation on 50 IEEE~118-bus scenarios (DeepSeek-V3, 2026-02-23) achieved 84.0\% tool-selection accuracy and 100\% parsing accuracy. A separate 56-scenario self-correction suite passed 49 of 56 cases (87.5\%) with a mean score of 89.3. These results establish a reproducible baseline for continued refinement while maintaining auditable, simulation-grounded decision support.
\end{abstract}


\section*{Nomenclature}
\begin{center}
\begin{tabular}{p{0.16\linewidth}p{0.74\linewidth}}
CIA     & Connection Impact Assessment \\
IBR     & Inverter-Based Resource \\
LLM     & Large Language Model \\
NL      & Natural Language \\
SCR     & Short-Circuit Ratio \\
ABC     & Abstract Base Class \\
OPF     & Optimal Power Flow \\
PF      & Power Flow \\
EMT     & Electromagnetic Transient \\
$\mathcal{T}$   & Tool registry: set of callable functions \\
$\mathcal{L}$   & Lessons learned (persistent memory) \\
$\mathcal{M}$   & Study memory (persistent structured store) \\
$f_k$           & Fidelity level $k \in \{1,2,3,4\}$ \\
$\mathcal{V}$   & Violation report from inspector \\
\end{tabular}
\end{center}

\section{Introduction}

The generator interconnection queue has emerged as a critical bottleneck in the
clean energy transition. As of 2023, over 2,600~GW of generation and storage
capacity awaited processing in U.S. regional transmission organization
queues~\cite{Rand_2024}, with mean study durations increasing from 2.1 to 5.0
years and withdrawal rates exceeding 80\% in certain regions~\cite{Johnston_2023}.
Although FERC Order No.~2023 mandates accelerated processing timelines and cluster-based
studies~\cite{ferc_order_2023}, the fundamental engineering
bottleneck---Connection Impact Assessment (CIA)---persists as a predominantly manual,
labor-intensive process requiring sequential power flow, contingency,
transient stability, and electromagnetic transient analyses.

Recent advances in LLM-based agents have demonstrated that language models
can autonomously employ tools to solve complex tasks. ReAct~\cite{ReAct_2023}
and Toolformer~\cite{Toolformer_2023} established that LLMs can learn to invoke
APIs through function calling, while ToolLLM~\cite{ToolLLM_2024} scaled
this paradigm to over 16,000 real-world APIs. In software engineering,
SWE-Agent~\cite{SWEAgent_2024} autonomously resolves GitHub issues with
minimal human intervention.
Open-source frameworks such as OpenClaw~\cite{OpenClaw_2025} further demonstrate
production-grade agent architectures featuring model-agnostic gateways,
skill plugins, and persistent memory---design patterns that are directly
transferable to domain-specific applications.

Despite this progress, LLM agents for power system operations remain in
an early stage of development. Prior work has explored LLMs for power system
analysis~\cite{PowerGPT_2024}, grid operator co-pilots~\cite{ChatGrid_2024},
and foundation models for grid intelligence~\cite{GridIntelligence_2024};
however, no existing system bridges the gap between natural-language interaction
and actual multi-fidelity simulation execution. Current approaches
either generate analysis code without executing it or operate on
simplified models that are disconnected from production-grade solvers.

This paper presents Grid-Mind, an LLM-orchestrated agent that interprets natural-language interconnection requests and extracts structured parameters---bus location, capacity, and resource type---without relying on rigid rule-based parsing. The agent dispatches multi-fidelity simulations through an LLM-first architecture, leveraging an eleven-tool registry and a solver-agnostic base class that supports established simulation engines including PandaPower~\cite{Thurner_2018}, ANDES~\cite{Cui_2021}, ParaEMT~\cite{ParaEMT_2023}, and PSS/E~\cite{PSSE_2024}. This architecture enables the language model to autonomously plan multi-step workflows, chaining up to five tool invocations within a single conversational turn. Critically, Grid-Mind grounds every approval or rejection in quantitative violation inspections that employ screening-level criteria informed by NERC~TPL standards~\cite{NERC_TPL_2023}. A three-layer anti-hallucination defense mitigates numerical fabrication risk by routing high-risk quantitative queries through deterministic tools and appending grounding warnings when warranted. To support continuous operation, the system maintains persistent memory of study results across sessions for auditability, while progressively refining its capabilities through a prompt-level lesson-optimization feedback loop---distinct from gradient-based policy learning---that distills operational failures into persistent lessons.

The benchmark harness supports five frontier LLMs accessed via
OpenRouter---Claude~3.5 Sonnet, GPT-4o, DeepSeek-R1~\cite{DeepSeek_R1_2025}, DeepSeek-V3,
and Qwen~2.5---evaluated on 50 diverse scenarios encompassing complete requests,
ambiguous multi-turn conversations, and edge cases. In this revision, we report
reproducible end-to-end results from the full agent loop for DeepSeek-V3 along with
the latest self-correction regression results from the current codebase.
The architecture parallels production agent frameworks such as OpenClaw's
gateway--skills--memory pattern~\cite{OpenClaw_2025}, while specializing it for power system
domain knowledge and physics-grounded validation.

\section{Related Work}

\subsection{LLM Agents and Tool Calling}
The emergence of function-calling capabilities in LLMs has enabled
autonomous tool use across diverse domains. ReAct~\cite{ReAct_2023} interleaves reasoning
traces with action execution, while Toolformer~\cite{Toolformer_2023}
demonstrates self-supervised API usage acquisition from demonstrations.
OpenAI's function-calling protocol~\cite{OpenAI_FunctionCalling_2023}
formalized tool specifications as JSON schemas, establishing what has become a
de facto industry standard. ToolLLM~\cite{ToolLLM_2024} subsequently extended this paradigm to
over 16,000 APIs with automated tool selection.

Production-grade agent frameworks have matured rapidly in recent years.
OpenClaw~\cite{OpenClaw_2025} implements a gateway--skills--memory
architecture in which a model-agnostic runtime orchestrates tool plugins
with persistent state, while SWE-Agent~\cite{SWEAgent_2024} specializes
this pattern for software engineering tasks.
These frameworks collectively demonstrate that LLM agents can reliably orchestrate
complex multi-step workflows when appropriately grounded in tool outputs.

\subsection{AI for Power System Operations}
Machine learning techniques have been extensively applied to power system stability
assessment, optimal power flow approximation, and load
forecasting~\cite{GridIntelligence_2024}. More recently, several studies have begun
exploring LLMs specifically for grid applications:
\cite{PowerGPT_2024} surveys LLM applications in power system
analysis, while \cite{ChatGrid_2024} proposes a ChatGPT-powered
grid operator co-pilot concept. However, these approaches
primarily leverage LLMs for text generation and code assistance
rather than for autonomous simulation orchestration.

\subsection{Interconnection Study Automation}
CIM standards~\cite{CIM_Middleware_2003}, graph
databases~\cite{Ravikumar_2017}, ontologies~\cite{Ma_Qian},
and digital twins~\cite{Bai_2022} have been proposed for grid
modeling, while hosting capacity studies~\cite{Glista_2024} inform
siting decisions. Nevertheless, none of these approaches integrate
a natural-language interface with multi-fidelity simulation
execution.

Table~\ref{tab:comparison} identifies the salient gap: among the systems
reviewed, no prior work combines an LLM agent with real solver execution,
multi-fidelity cascading, and physics-grounded violation checking.

\begin{table}[t]
\caption{Feature Comparison: Prior Work vs. Grid-Mind}
\label{tab:comparison}
\centering\small
\begin{tabular}{lcccccc}
\toprule
Approach & NL & Solver & Multi-F. & Viol. & Mem. & Self-Corr. \\
\midrule
ReAct~\cite{ReAct_2023}       & \checkmark & --         & --         & --         & --         & -- \\
ToolLLM~\cite{ToolLLM_2024}   & \checkmark & --         & --         & --         & --         & -- \\
OpenClaw~\cite{OpenClaw_2025}  & \checkmark & --         & --         & --         & \checkmark & -- \\
PowerGPT~\cite{PowerGPT_2024} & \checkmark & --         & --         & --         & --         & -- \\
ChatGrid~\cite{ChatGrid_2024}  & \checkmark & partial    & --         & --         & --         & -- \\
Digital Twin~\cite{Bai_2022}   & --         & \checkmark & --         & --         & --         & -- \\
\textbf{Grid-Mind}             & \checkmark & \checkmark & \checkmark & \checkmark & \checkmark & \checkmark \\
\bottomrule
\end{tabular}
\end{table}

\section{System Architecture}

\begin{figure*}[t]
\centering
\begin{tikzpicture}[
    node distance=0.5cm and 0.8cm,
    layer/.style={rectangle, rounded corners=4pt, draw=gray!40, thick, minimum width=15.5cm, minimum height=1.5cm, font=\small, drop shadow={opacity=0.03, shadow xshift=2pt, shadow yshift=-2pt}},
    component/.style={rectangle, rounded corners=2pt, draw=gray!50, fill=#1, minimum height=0.85cm, font=\footnotesize, text centered, drop shadow={opacity=0.1, shadow xshift=1pt, shadow yshift=-1pt}},
    component/.default={blue!10},
    arrow/.style={-Stealth, thick},
    label/.style={font=\footnotesize\bfseries, text=#1},
]

\node[layer, fill=green!5] (ui_layer) {};
\node[label=green!60!black, anchor=west] at ([xshift=0.3cm]ui_layer.west) {Interface Layer};
\node[component=green!15, text width=2.5cm] (nl_input) at ([xshift=-3.5cm]ui_layer.center) {Natural Language\\Input};
\node[component=green!15, text width=2.5cm] (rest_api) at (ui_layer.center) {REST API\\(FastAPI)};
\node[component=green!15, text width=2.5cm] (dashboard) at ([xshift=3.5cm]ui_layer.center) {Web Dashboard};
\node[component=green!15, text width=1.8cm, font=\tiny] (heartbeat) at ([xshift=6.3cm]ui_layer.center) {Health\\Monitor};

\node[layer, fill=orange!5, below=0.6cm of ui_layer] (agent_layer) {};
\node[label=orange!70!black, anchor=west] at ([xshift=0.3cm]agent_layer.west) {Agent Layer};
\node[component=orange!15, text width=2.2cm] (tools) at ([xshift=-3.5cm]agent_layer.center) {Tool\\Registry};
\node[component=orange!15, text width=2.2cm] (agent) at (agent_layer.center) {Conversation\\Agent};
\node[component=orange!15, text width=2.2cm] (llm) at ([xshift=3.5cm]agent_layer.center) {LLM Backend\\(OpenRouter)};
\node[component=orange!15, text width=2.2cm] (lessons) at ([xshift=6.3cm]agent_layer.center) {Lessons\\+ Memory};

\node[layer, fill=blue!5, below=0.6cm of agent_layer] (sim_layer) {};
\node[label=blue!70!black, anchor=west] at ([xshift=0.3cm]sim_layer.west) {Simulation Layer};
\node[component=blue!15, text width=1.8cm] (ss) at ([xshift=-3.5cm]sim_layer.center) {Steady-State\\(AC PF)};
\node[component=blue!15, text width=1.8cm] (n1) at ([xshift=-1.0cm]sim_layer.center) {N-1\\Contingency};
\node[component=blue!15, text width=1.8cm] (ts) at ([xshift=1.5cm]sim_layer.center) {Transient\\Stability};
\node[component=blue!15, text width=1.8cm] (emt) at ([xshift=4.0cm]sim_layer.center) {EMT\\Screening};
\node[component=red!15, text width=1.8cm] (insp) at ([xshift=6.3cm]sim_layer.center) {Violation\\Inspector};

\node[layer, fill=purple!5, below=0.6cm of sim_layer] (solver_layer) {};
\node[label=purple!70!black, anchor=west] at ([xshift=0.3cm]solver_layer.west) {Solver Layer};
\node[component=purple!10, text width=2cm] (pp) at ([xshift=-2.25cm]solver_layer.center) {PandaPower\\Adapter};
\node[component=purple!10, text width=2cm] (andes) at ([xshift=1.5cm]solver_layer.center) {ANDES\\Adapter};
\node[component=purple!10, text width=2cm] (pemt) at ([xshift=4.0cm]solver_layer.center) {ParaEMT\\Adapter};
\node[component=purple!10, text width=2cm] (psse) at ([xshift=6.3cm]solver_layer.center) {PSS/E\\Adapter};

\node[font=\scriptsize\itshape, text=purple!60, below=0.1cm of solver_layer] {GridSolver Abstract Base Class};

\draw[arrow] (nl_input.south east) to[bend right=15] (agent.north west);
\draw[arrow] (rest_api) -- (agent);
\draw[arrow, dashed] (dashboard) -- (rest_api);
\draw[arrow, dashed, green!50!black] (heartbeat.north west) to[bend right=20] (rest_api.north east);

\draw[arrow, <->] (llm) -- (agent);
\draw[arrow, <->] (agent.south west) to[bend right=15] (tools.north east);
\draw[arrow, <->] (agent.north east) to[bend left=12] (lessons.north west);

\draw[arrow] (tools) -- (ss);
\draw[arrow, dashed] (ss) -- (n1);
\draw[arrow, dashed] (n1) -- (ts);
\draw[arrow, dashed] (ts) -- (emt);
\draw[arrow] (emt) -- (insp);
\draw[arrow] (ss.south east) to[bend right=15] (insp.south west);

\draw[arrow] (ss.south) -- (pp.north west);
\draw[arrow] (n1.south) -- (pp.north east);
\draw[arrow] (ts) -- (andes);
\draw[arrow] (emt) -- (pemt);

\draw[arrow, red!60, thick, dashed] (insp.east) -- ++(0.4,0) |- ([yshift=-0.1cm]lessons.south east) node[pos=0.6, right, font=\scriptsize, text=red!60, fill=white, inner sep=1pt] {Lesson Feedback};

\end{tikzpicture}
\caption{Grid-Mind architecture. The Agent Layer orchestrates multi-fidelity
simulations through a solver-agnostic tool registry. Persistent memory
stores study results across sessions. Dashed arrows indicate conditional
escalation. The prompt-level lesson feedback loop injects distilled
rules from failure analysis back into the agent's persistent memory. A health monitor provides
real-time system status.}
\label{fig:architecture}
\end{figure*}
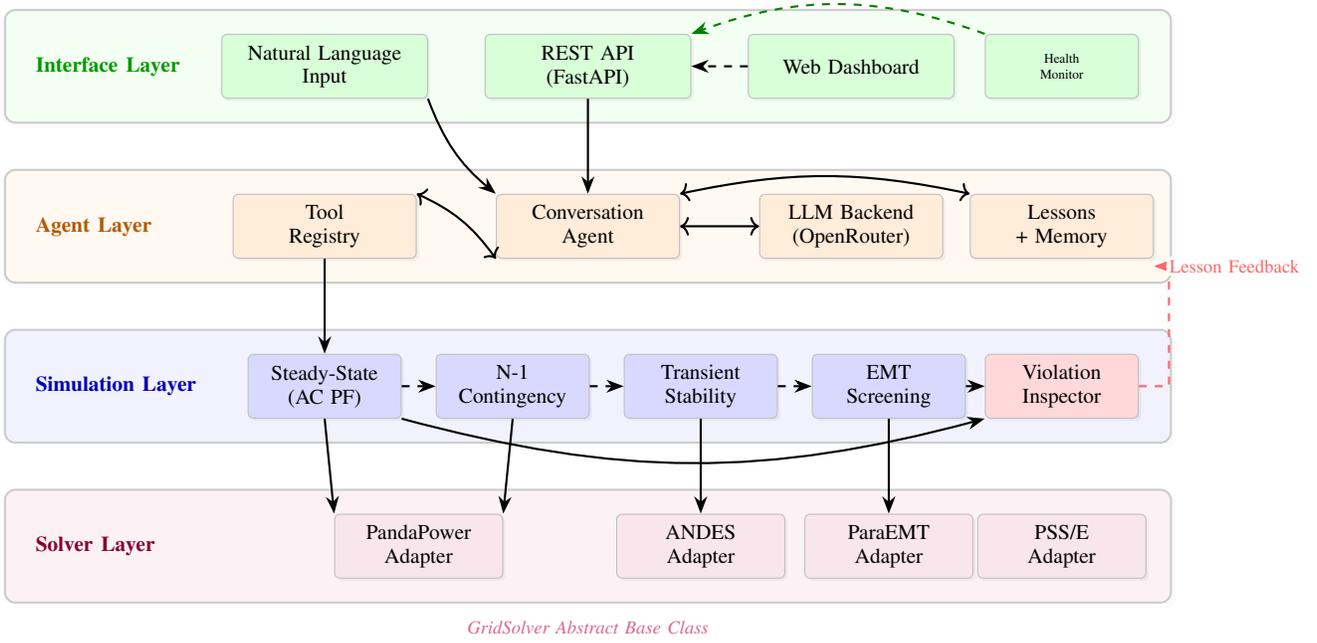

Grid-Mind adopts a layered architecture inspired by production agent
frameworks~\cite{OpenClaw_2025}, specialized for the requirements of the power system domain.
Fig.~\ref{fig:architecture} illustrates the four-layer design.

\subsection{Solver-Agnostic Abstract Base Class}

The \emph{GridSolver} Abstract Base Class (ABC) defines a uniform interface for power system solvers, comprising over 20 abstract methods that span case loading, power flow solution, bus and branch result access, contingency execution, and violation checking. Four concrete adapters implement this interface.

These adapters include PandaPower~\cite{Thurner_2018} for steady-state AC and optimal power flow computations essential to initial development and screening; ANDES~\cite{Cui_2021} for high-fidelity time-domain transient stability simulations required in inverter-based resource scenarios; ParaEMT~\cite{ParaEMT_2023} for rapid electromagnetic transient analysis in weak-grid environments with low short-circuit ratios; and a PSS/E~\cite{PSSE_2024} integration path for utility environments, which is license-dependent and currently implemented as a compatibility stub.

An adapter registry provides robust instantiation with lazy loading:
\begin{equation}
\begin{aligned}
\mathcal{S}(b) &\rightarrow \text{GridSolver}, \\
b &\in \{\text{PandaPower}, \text{ANDES}, \text{ParaEMT}, \text{PSS/E}\}
\end{aligned}
\label{eq:factory}
\end{equation}

This design enables the LLM agent to operate uniformly across backends,
with solver selection governed by the simulation layer rather than the agent itself.

\subsection{Violation Inspector}

The \emph{Violation Inspector} provides solver-agnostic violation detection against configurable screening criteria informed by NERC TPL-001 planning standards~\cite{NERC_TPL_2023}:

\begin{equation}
\begin{aligned}
\mathcal{V}(s) &= \{v \mid v \in \text{check}(s, \ell)\} \\
&\quad \text{where } \ell = (V_{\min}, V_{\max}, L_{\max}, \delta_{\max})
\end{aligned}
\label{eq:violations}
\end{equation}

where $s$ denotes a solver result, $\ell$ represents a limit configuration, and each
violation $v$ encapsulates the element type, index, violation type, observed value,
applicable limit, and margin percentage. The inspector distinguishes \emph{hard}
violations from \emph{borderline} conditions within configurable tolerance
bands ($\pm 0.01$~p.u.\ for voltage, $\pm 5\%$ for loading).
In the current implementation, this inspector governs steady-state and
contingency acceptance, while transient and EMT acceptance employ stage-specific
criteria summarized in Table~\ref{tab:stage_criteria}.

\section{Multi-Fidelity CIA Pipeline}

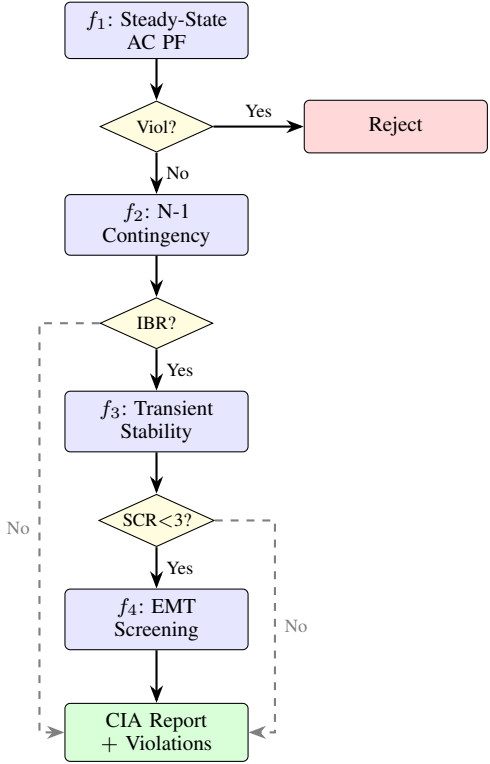
\begin{figure}[t]
\centering
\begin{tikzpicture}[node distance=0.55cm,
    stage/.style={rectangle, rounded corners=2pt, draw, fill=blue!10,
      text width=2.2cm, text centered, minimum height=0.7cm, font=\footnotesize},
    decision/.style={diamond, draw, fill=yellow!15, text width=0.9cm,
      text centered, aspect=2.2, inner sep=1pt, font=\scriptsize},
    arrow/.style={-Stealth, thick},
    skip/.style={-Stealth, thick, dashed, gray}
]
\node[stage] (ss) {$f_1$: Steady-State\\AC PF};
\node[decision, below=of ss] (d1) {Viol?};
\node[stage, below=of d1] (n1) {$f_2$: N-1\\Contingency};
\node[decision, below=of n1] (d2) {IBR?};
\node[stage, below=of d2] (ts) {$f_3$: Transient\\Stability};
\node[decision, below=of ts] (d3) {SCR$<$3?};
\node[stage, below=of d3] (emt) {$f_4$: EMT\\Screening};
\node[stage, below=0.7cm of emt, fill=green!15] (report) {CIA Report\\$+$ Violations};

\node[stage, right=1.2cm of d1, fill=red!15] (rej) {Reject};

\draw[arrow] (ss) -- (d1);
\draw[arrow] (d1) -- node[right, font=\scriptsize] {No} (n1);
\draw[arrow] (d1) -- node[above, font=\scriptsize] {Yes} (rej);
\draw[arrow] (n1) -- (d2);
\draw[arrow] (d2) -- node[right, font=\scriptsize] {Yes} (ts);
\draw[skip] (d2.west) -- ++(-0.8,0) |- (report.west) node[pos=0.25, left, font=\scriptsize] {No};
\draw[arrow] (ts) -- (d3);
\draw[arrow] (d3) -- node[right, font=\scriptsize] {Yes} (emt);
\draw[skip] (d3.east) -- ++(0.8,0) |- (report.east) node[pos=0.25, right, font=\scriptsize] {No};
\draw[arrow] (emt) -- (report);

\end{tikzpicture}
\caption{Multi-fidelity CIA pipeline. Stages are conditionally activated
based on connection properties and violation severity.}
\label{fig:pipeline}
\end{figure}

The pipeline orchestrates a four-stage assessment cascade (Fig.~\ref{fig:pipeline}), in which each successive stage provides progressively higher fidelity analysis.

The assessment commences with a steady-state AC power flow ($f_1$) to evaluate the immediate voltage and thermal impacts of the proposed connection. Provided that the base topology remains secure, the pipeline automatically advances to N-1 contingency analysis ($f_2$), performing systematic equipment outage screening against emergency thermal limits. Conditional escalation into dynamic analysis follows: if the request involves an inverter-based resource ($\text{IBR} = \text{true}$), the pipeline may invoke a time-domain transient stability simulation ($f_3$). When EMT analysis is enabled for IBR requests, an EMT screening stage ($f_4$) is executed, which evaluates the short-circuit ratio against a configurable threshold (default 3.0).

The escalation policy is formally expressed as:
\begin{equation}
f^*(r) = \max\left\{k : \phi_k(r) = \text{true}\right\}
\label{eq:fidelity}
\end{equation}
where $\phi_k$ are predicate functions encoding escalation conditions:
$\phi_1 = \text{true}$ (always), $\phi_2 = e_c$ (contingency enabled),
$\phi_3 = e_t \wedge \text{IBR}(r)$, $\phi_4 = e_e \wedge \text{IBR}(r)$,
where $e_c$, $e_t$, $e_e$ denote Boolean flags for contingency, transient, and EMT enablement respectively.

The implementation-level acceptance criteria are detailed below:
\begin{table}[t]
\caption{Stage Acceptance Criteria (Current Implementation)}
\label{tab:stage_criteria}
\centering\small
\begin{tabular}{p{1.45cm}p{5.7cm}}
\toprule
Stage & Pass/Fail Criterion \\
\midrule
Steady-state ($f_1$) &
PF converges and no hard violations under NORMAL limits
(voltage 0.95--1.05 p.u., thermal Rate A 100\%);
angle-difference screening (30$^\circ$) is optional and disabled by default;
borderline if within 0.01 p.u. of voltage limits or within 5\% of thermal limits, with optional OPF escalation. \\
\midrule
N-1 ($f_2$) &
EMERGENCY limits (voltage 0.90--1.10 p.u., thermal Rate B 110\%);
fails when the project introduces new post-contingency failures relative to baseline N-1, and can optionally fail on material worsening of pre-existing failed contingencies (default threshold: +2.0\% max-margin erosion). \\
\midrule
Transient ($f_3$) &
ANDES run to configured horizon (default 10 s) with configurable fault
(default 3-phase at POI, 0.1--0.2 s); pass requires completed run and no loss-of-synchronism heuristic trigger (final generator-angle spread $\leq 2\pi$ rad). \\
\midrule
EMT ($f_4$) &
ParaEMT SCR screen for IBR buses; fails if any screened bus has
$\mathrm{SCR}_b < \mathrm{SCR}_{\min}$ (default $\mathrm{SCR}_{\min}=3.0$). \\
\bottomrule
\end{tabular}
\end{table}

For EMT screening, SCR is defined as
\begin{equation}
\mathrm{SCR}_b = \frac{S_{\mathrm{sc},b}}{S_{\mathrm{IBR},b}}.
\label{eq:scr}
\end{equation}
The current adapter estimates $S_{\mathrm{sc},b}$ from the diagonal Y-bus
admittance when exposed by the backend,
$S_{\mathrm{sc},b} \approx |Y_{bb}|\,S_{\mathrm{base}}$ with
$S_{\mathrm{base}} = 100$~MVA and $|V_b| \approx 1$~p.u.; when this quantity is unavailable, the system
employs a conservative system-size heuristic as a fallback. This constitutes a screening-level
proxy and should be superseded by utility-specific short-circuit studies for
production-grade approvals.

Regarding N-1 policy, the default fail-on-new-failures setting targets incremental
impact screening---specifically, project-caused reliability degradation---across heterogeneous test
cases. For more stringent planning practice, the implementation provides an opt-in
material-worsening failure mode (Table~\ref{tab:stage_criteria}) that flags
meaningful erosion of pre-existing contingency margins.

The pipeline produces a structured interconnection impact report comprising
per-stage results, violation details, and a final recommendation
(approve, reject, or borderline) accompanied by reason codes. This report serves as the
factual basis upon which the LLM constructs its natural-language explanation.

\subsection{Binary-Search Capacity Tool}

Beyond pass/fail assessment, Grid-Mind provides a
binary capacity search operator that determines the maximum active power
a bus can accept before violations occur. This tool performs a
bisection search over the MW range $[\text{min}, \text{max}]$:
at each iteration, a full CIA is executed at the midpoint, the search interval is
narrowed based on the resulting approval status, and the procedure terminates when the interval
width falls below a configurable tolerance (default: 1~MW).
In practice, 8--9 iterations suffice for a 0--500~MW range. Results
are automatically persisted to the memory system
(Section~\ref{sec:memory}) for subsequent retrieval.

Because bisection assumes monotone feasibility with respect to injected power,
the implementation incorporates a monotonicity contradiction check
(e.g., approval at a higher MW level following a lower-MW rejection). Upon detection,
the operator records explicit diagnostics and reverts to a coarse
range scan, reporting the highest sampled feasible point rather than
asserting a strict bisection boundary.

At the rejection boundary, the tool enriches its output with a
structured \emph{rejection explanation}: the limiting factor
(e.g., steady-state violation, contingency failure, convergence
divergence), the failing assessment stage(s), and any
project-caused violations with element, type, value, and limit.
This enables the agent to explain \emph{why} the capacity limit
was reached, not just the numerical boundary.

\section{LLM Agent Design}

\subsection{Agent Architecture}

The conversational agent (Algorithm~\ref{alg:agent}) implements
an \textbf{LLM-first} design philosophy. In contrast to conventional
agent architectures where deterministic routing handles the majority of requests and the
LLM serves as a fallback, Grid-Mind positions the LLM at the center of
decision-making, augmented by two pre-LLM safety guardrails:
(i)~anti-hallucination capacity routing and
(ii)~required-input clarification for CIA-like prompts.

Given a user message history $H = [m_1, \ldots, m_n]$, the agent:

\begin{enumerate}
\item Checks forced capacity routing (anti-hallucination guardrail);
\item Checks required-input clarification for CIA-like prompts and returns
a deterministic clarification prompt when mandatory fields are missing;
\item Builds a system prompt incorporating a \textbf{planning and
reflection} framework, lessons $\mathcal{L}$, relevant memory entries
from $\mathcal{M}$, and auto-detected context hints (parameters
extracted from conversation, last report status, detected mitigations);
\item Submits $[s] \cup H$ to the LLM with all eleven tool specifications
$\mathcal{T}$;
\item The LLM \textbf{plans} its approach, chains up to 5 tool-call
rounds (e.g., run OPF $\rightarrow$ check remaining violations
$\rightarrow$ suggest mitigations), and \textbf{reflects} on each
result before deciding the next action;
\item Validates the final response for ungrounded numerical claims,
persists results to memory, and returns.
\end{enumerate}

Planning and reflection behavior is induced through the foundational identity prompt, which instructs the LLM to reason through four considerations: what the user is requesting, what data is required, the appropriate sequence of tool invocations, and whether each intermediate result fully addresses the question. Context hints pre-populate detected parameters (bus, MW, type, case, mitigations) so the LLM can make informed decisions without expending a round-trip on parsing.

\begin{algorithm}[t]
\caption{LLM-First Agent Loop with Anti-Hallucination}
\label{alg:agent}
\begin{algorithmic}[1]
\Require History $H$, tools $\mathcal{T}$, lessons $\mathcal{L}$, memory $\mathcal{M}$
\Ensure Response text $y$, report $R$
\If{IsCapacityQuestion($H$)} \Comment{Safety guardrail}
  \If{MissingRequiredCapacityInputs($H$)}
    \State \Return ClarificationPrompt, $\varnothing$
  \EndIf
  \State $R \leftarrow \mathcal{T}.\text{execute}(\text{find\_max\_capacity})$
  \State \Return $\text{Summarize}(R), R$
\EndIf
\If{MissingRequiredCIAInputs($H$)} \Comment{Safety guardrail}
  \State \Return ClarificationPrompt, $\varnothing$
\EndIf
\State $s \leftarrow \text{BuildPrompt}(\mathcal{L}, \mathcal{M}, H)$
\State $s \leftarrow s \oplus \text{ContextHints}(H)$ \Comment{Pre-extracted params}
\State $\text{msgs} \leftarrow [s] \cup H$
\State $\text{tool\_called} \leftarrow \text{false}$
\For{$i = 1$ \textbf{to} $5$} \Comment{Multi-step tool chaining}
  \State $\text{resp} \leftarrow \text{LLM.chat}(\text{msgs}, \mathcal{T})$
  \If{resp has tool\_calls}
    \ForAll{call $c$ in resp.tool\_calls}
      \State $R_c \leftarrow \mathcal{T}.\text{execute}(c.\text{name}, c.\text{args})$
      \State Append result to msgs; $\text{tool\_called} \leftarrow \text{true}$
      \If{$R_c$ has report} $R \leftarrow R_c$ \EndIf
    \EndFor
  \Else
    \State $y \leftarrow \text{resp.content}$; \textbf{break}
  \EndIf
\EndFor
\If{HasUngroundedNumerics($y$, tool\_called)} \Comment{Layer 3}
  \State $y \leftarrow y \oplus \text{GroundingWarning}$
\EndIf
\State SaveToMemory($R, \mathcal{M}$)
\State \Return $y, R$
\end{algorithmic}
\end{algorithm}

\subsection{Action Space}

The \emph{Action Registry} exposes simulation capabilities as OpenAI-format function specifications~\cite{OpenAI_FunctionCalling_2023}.
The registry comprises eleven tools spanning assessment, analysis,
topology queries, and system management.
Table~\ref{tab:tool_specs} enumerates the complete interface contract,
including required arguments, key return payloads, and built-in safeguards.

\begin{table*}[t]
\caption{Tool Interface Summary (Current Implementation)}
\label{tab:tool_specs}
\centering\scriptsize
\setlength{\tabcolsep}{4pt}
\begin{tabular}{p{3.0cm}p{3.6cm}p{3.2cm}p{6.0cm}}
\toprule
Tool & Required Arguments & Key Returns & Safeguards / Notes \\
\midrule
\texttt{list\_backends},\newline \texttt{list\_cases} & none &
available backends / test cases &
Read-only metadata queries. \\
\addlinespace[2pt]
\texttt{set\_backend} & \texttt{backend} &
success flag, active backend &
Name validated against registered adapters. \\
\addlinespace[2pt]
\texttt{run\_powerflow} & \texttt{case\_path} &
convergence flag, bus/branch results, violations &
Convergence status surfaced explicitly; numerics from solver outputs. \\
\addlinespace[2pt]
\texttt{run\_opf} & \texttt{case\_path} &
convergence flag, post-OPF violations, dispatch &
No topology mutation; includes post-OPF violation inspection. \\
\addlinespace[2pt]
\texttt{inspect\_violations} & \texttt{case\_path} &
structured violation report &
Angle-difference check configurable; disabled by default. \\
\addlinespace[2pt]
\texttt{run\_contingency} & \texttt{case\_path} &
N-1 pass/fail counts, failing contingencies &
Runs baseline PF first; explicit error on non-convergence. \\
\addlinespace[2pt]
\texttt{run\_cia} & \texttt{case\_path},\newline \texttt{connection} &
stage reports, final decision, reason codes &
Missing-field clarification gate; baseline-aware N-1 impact logic. \\
\addlinespace[2pt]
\texttt{run\_cia\_with\_\-mitigation} &
\texttt{case\_path},\newline \texttt{connection},\newline \texttt{mitigations} &
CIA report + applied mitigations &
Mitigations constrained to shunt-like interventions for traceability. \\
\addlinespace[2pt]
\texttt{find\_max\_capacity} &
\texttt{case\_path},\newline \texttt{bus},\newline \texttt{connection\_type} &
max approved MW, boundary reports, diagnostics &
Forced routing for capacity queries; monotonicity contradiction detection. \\
\addlinespace[2pt]
\texttt{query\_network\_data} & \texttt{case\_path} &
raw topology / equipment data &
Read-only introspection; no solved operating point claimed. \\
\bottomrule
\end{tabular}
\end{table*}

Conceptually, the registry organizes capabilities into four groups.
\emph{Assessment operators} execute the full Connection Impact Assessment cascade, with options to evaluate base configurations or to pre-install reactive compensation mitigations.
\emph{Analysis operators} provide granular access to the underlying physics, enabling the agent to request steady-state AC power flows for per-bus voltage examination, invoke optimal power flow (OPF) routines for redispatch scheduling, perform comprehensive voltage and thermal violation scans, and execute systematic N-1 contingency screening.
\emph{Capacity search operators} employ binary-search routines to determine the maximum allowable generation or load injection at a specified bus before violations occur. Finally, \emph{topology and system operators} permit the agent to extract detailed grid parameters (e.g., branch impedances and generator limits) and manage simulation cases and solver backends.

All tool specifications adhere to the JSON Schema format, enabling invocation by any
OpenAI-compatible LLM. The primary assessment operator accepts nested parameters reflecting the structured nature of interconnection requests, including network bus, capacity, resource type, synchronization status, and flags for contingency or transient screening. This model-agnostic design mirrors the ``skills and plugins'' pattern of production agent frameworks~\cite{OpenClaw_2025}, wherein capabilities are defined declaratively and can be extended without modifying the agent core.

\subsection{Natural Language Parsing}

A critical function of the agent is the extraction of structured
interconnection request parameters from free-form natural language.
The extraction targets four fields:

\begin{equation}
\text{parse}(m) \rightarrow (b, P, \tau, \iota) \quad
b \in \mathbb{Z}^+, \; P \in \mathbb{R}^+, \; \tau \in \mathcal{C}, \; \iota \in \{0,1\}
\label{eq:parse}
\end{equation}

where $b$ denotes the bus number, $P$ represents active power in MW, $\tau$ is the connection
type drawn from the set $\mathcal{C} = \{\text{load}, \text{solar}, \text{wind},
\text{bess}, \text{hybrid}, \text{synchronous}\}$, and $\iota$ indicates
IBR status. When any parameter is absent, the agent solicits
clarification rather than inferring a value. In the current
implementation, a conservative clarification policy is enforced for missing
resource type---no default ``load'' fallback is applied unless the request explicitly
contains a load-indicative domain term (e.g., \emph{data center}). The same
policy governs the direct capacity-search routing path: if
the resource type remains unresolved after context lookup, the
agent requests clarification before invoking the capacity-search operator.

\subsection{Physics Grounding with Engineering Judgment}

A fundamental design principle is that quantitative claims must be
grounded in simulation output whenever tool data is available. The LLM generates natural-language
explanations, but approval and rejection decisions together with underlying violation data originate
exclusively from the solver and inspector.

The LLM is nonetheless \emph{encouraged} to provide \textbf{engineering judgment} on tool results: characterizing margin severity (``a $-2.1\%$ margin is relatively mild''), contextualizing capacity limits (``50~MW is substantial for this bus''), and recommending multi-step mitigation workflows. This separation of concerns---\emph{specific numerical values} must originate from simulation operators, whereas \emph{qualitative interpretation} leverages the LLM's domain knowledge---is encoded explicitly in both the architectural identity instructions and the operational system prompt.

\subsection{Anti-Hallucination Defense-in-Depth}
\label{sec:anti_hallucination}

LLM hallucination---the generation of plausible but factually incorrect content---is a well-documented challenge~\cite{Huang_Hallucination_2025, Tonmoy_Hallucination_2024}. Despite physics grounding of simulation outputs, we observed that reasoning-trained LLMs (notably DeepSeek-R1) occasionally fabricate \emph{numerical answers to quantitative questions}, bypassing the tool-calling path entirely. For instance, when queried for the maximum load at bus~14 on the IEEE 118-bus system, the LLM responded with a fabricated value of approximately 127~MW without invoking the capacity-search operator; the verified limit was 3.9~MW---a $33\times$ discrepancy. This failure mode is addressed through three complementary defense layers:

\textbf{Layer~1: System instruction hardening.} Both the agent's foundational identity instructions and operational prompt contain explicit anti-fabrication directives: \emph{never state specific MW, pu, MVA, or percentage values for individual grid elements unless those values originated from a physics operator in the current conversation or constitute well-known published standards}.
The prompt additionally incorporates \emph{memory usage rules} that prevent the
agent from misrepresenting session-local memory entries as independent
historical data (e.g., stating ``historical studies confirm\ldots''
when citing a result from an earlier simulation in the same session).
This constitutes a soft guardrail that depends on instruction-following fidelity.

\textbf{Layer~2: Forced action routing.} A deterministic pre-LLM classifier identifies two high-risk families of capacity queries: (i)~specific-bus capacity questions combining capacity intent with an explicit bus reference, accommodating both resource/power wording and generic forms such as ``max capacity at bus~14''; and (ii)~``best bus for maximum capacity'' questions combining capacity intent with a best-bus intent phrase. When triggered, the system directly invokes the binary-search capacity operator, bypassing the LLM entirely for that request class. If required inputs (e.g., resource type) remain unspecified, the system solicits clarification prior to execution.

\textbf{Layer~3: Post-response grounding validator.} After the LLM
generates a response, a regex-based scanner examines the output for ungrounded
numerical claims (e.g., patterns of the form ``$X$~MW,'' ``$X$~pu,''
or ``capacity is~$X$''). Each match is evaluated against an
150-character context window for \emph{safe phrases} (e.g., NERC standard values,
per-unit definitions). If no grounding-capable tool was invoked during the turn and an
ungrounded numeric pattern is detected, a disclaimer is appended
directing the user to request a simulation. This mechanism currently operates as a
turn-level detector rather than per-number provenance tracing. To mitigate false
grounding credit, invocations of non-analytical tools (e.g., backend listing or
switching) do not exempt a response from Layer~3 scrutiny.

Table~\ref{tab:hallucination_defense} summarizes the defense layers
and their properties.

\begin{table}[t]
\caption{Anti-Hallucination Defense Layers}
\label{tab:hallucination_defense}
\centering\small
\begin{tabular}{lccc}
\toprule
Layer & Type & Bypass-proof & Coverage \\
\midrule
Prompt hardening   & Soft    & No  & All queries \\
Forced routing     & Hard    & Yes & Capacity Qs \\
Grounding validator & Detect & No  & No-tool responses \\
\bottomrule
\end{tabular}
\end{table}

\subsection{Persistent Memory}
\label{sec:memory}

Inspired by the persistent state patterns of production agent
frameworks~\cite{OpenClaw_2025}, Grid-Mind maintains an append-only structured
memory system that stores completed CIA studies and capacity search
results across sessions. Memory injection prioritizes current-conversation
context, while retrieval can access recent global memory entries.
Each study record captures the timestamp,
test case, bus, MW, connection type, approval status, violation counts,
and a human-readable summary.

The memory system supports four recall modes: (i)~bus-specific recall
for retrieving past studies at a given bus and case, (ii)~case-wide
recall for browsing all studies on a network, (iii)~keyword search
across summaries, and (iv)~max-capacity recall for retrieving previously
computed hosting limits. On each LLM invocation, relevant memory
entries are injected into the system prompt, enabling the agent to
reference prior results (e.g., ``the last study at bus~14 identified a
3.9~MW limit'') without re-executing simulations.
Critically, the memory injection includes an explicit caveat that
these entries are from \emph{earlier simulations in the current
session}, not independent historical data.  This prevents a failure
mode we observed where the LLM presented session-local results as
authoritative ``historical studies'' or ``past analyses,'' lending
false credibility to what were merely prior runs in the same
conversation.  The agent is instructed to prefer fresh simulations
for new questions and to cite memory only as supplementary context.

A human-readable study ledger is automatically regenerated upon each memory insertion, providing a transparent audit trail suitable for commitment to regulatory version control systems.

\section{Self-Improving Prompt-Lesson Loop}

\begin{figure}[t]
\centering
\begin{tikzpicture}[node distance=0.7cm,
    box/.style={rectangle, rounded corners=2pt, draw, fill=blue!8,
      text width=2.5cm, text centered, minimum height=0.7cm, font=\footnotesize},
    arrow/.style={-Stealth, thick}
]
\node[box] (gen) {Scenario\\Generator};
\node[box, right=of gen] (eval) {Evaluator\\(Score)};
\node[box, below=of eval] (opt) {Optimizer\\(LLM Analysis)};
\node[box, left=of opt] (mem) {Lessons\\Learned};

\draw[arrow] (gen) -- (eval);
\draw[arrow] (eval) -- (opt);
\draw[arrow] (opt) -- (mem);
\draw[arrow, dashed] (mem) -- (gen);
\draw[arrow, red!60, thick] (mem.north) -- (gen.south) node[midway, left, font=\scriptsize, text=red!60] {Inject};
\end{tikzpicture}
\caption{Self-improving prompt-level lesson loop. Lessons from failure analysis
are injected into the system prompt for subsequent iterations.}
\label{fig:rl_loop}
\end{figure}
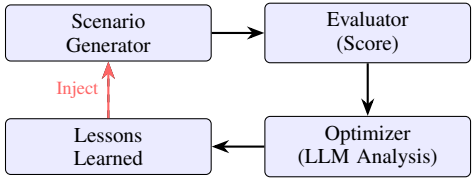

Grid-Mind incorporates a lightweight prompt-level self-correction mechanism
(Fig.~\ref{fig:rl_loop}) that improves agent performance without
requiring model retraining.

\subsection{Scenario Generation}

The dataset generator produces multi-turn conversation scenarios
covering four categories: (i)~ambiguous requests requiring clarification,
(ii)~partial requests with missing parameters, (iii)~complete
requests spanning all connection types, and (iv)~follow-up
questions pertaining to results and criteria.

\subsection{Evaluation}

Each agent response is scored across three weighted dimensions:
\begin{equation}
S = w_a \cdot S_{\text{action}} + w_c \cdot S_{\text{content}} + w_f \cdot S_{\text{format}}
\label{eq:score}
\end{equation}
with $w_a = 0.5$, $w_c = 0.35$, $w_f = 0.15$.
Action scoring checks tool selection correctness; content scoring
verifies keyword presence and factual accuracy; format scoring
assesses response structure.

\subsection{Optimization}

The optimizer analyzes failed scenarios---those scoring below a
configurable threshold---using an LLM to generate concise, actionable lessons.
These lessons are appended to a persistent repository and injected into the system prompt for all subsequent sessions, thereby completing the self-improvement loop.
\begin{equation}
s_{t+1} = s_0 \oplus \mathcal{L}_t
\label{eq:lessons}
\end{equation}

where $s_0$ is the base system prompt and $\mathcal{L}_t$ denotes the
accumulated lesson set at iteration $t$. It is important to note that this is \emph{not} reinforcement
learning in the control-theoretic or policy-gradient sense; rather, it is
prompt-level context optimization. The approach shares with RLHF~\cite{RLHF_2022}
only the high-level objective of iterative behavioral improvement, while
remaining applicable to closed-source API-accessed models.

\section{Experimental Setup}

\subsection{Test Systems}

The fixed benchmark employs the IEEE 118-bus test case loaded
from PandaPower's built-in network library~\cite{Thurner_2018}. This system
comprises 118 buses, 186 branches, and 54 generators, providing
sufficient complexity for realistic interconnection study scenarios.

The self-correction regression loop operates on a mixed-case suite
(IEEE~14/30/57/118), reflecting the broader operational scope
supported by the current agent implementation.

\subsection{Benchmark and Regression Scenarios}

The benchmark comprises 50 scenarios distributed across nine categories
(Table~\ref{tab:scenarios}):

\begin{table}[t]
\caption{Benchmark Scenario Categories}
\label{tab:scenarios}
\centering\small
\begin{tabular}{lcp{4.2cm}}
\toprule
Category & $N$ & Description \\
\midrule
Complete load       & 8  & All parameters specified (load) \\
Complete generation & 8  & All parameters specified (solar/wind/BESS) \\
Missing bus         & 5  & Bus number omitted \\
Missing MW          & 5  & Power capacity omitted \\
Missing type        & 5  & Connection type omitted \\
Multi-turn          & 6  & Clarification then execution \\
Follow-up           & 5  & Questions about prior results \\
Edge cases          & 4  & Invalid bus, zero MW, huge load \\
Theory questions    & 4  & Conceptual questions (no CIA needed) \\
\midrule
\textbf{Total}      & \textbf{50} & \\
\bottomrule
\end{tabular}
\end{table}

For ongoing agent improvement, the prompt-level self-correction loop evaluates a separate
56-scenario regression suite encompassing mitigation-realism prompts
(OPF versus manual setpoint/tap edits), max-capacity follow-ups, and
contingency/impact-consistency checks.
To facilitate reproducibility, the scenario definitions,
benchmark runner, and timestamped result artifacts used in the reported tables
are included in the accompanying repository.\footnote{Scenario definitions, benchmark runner, and timestamped result artifacts are provided to enable full reproducibility.}

\subsection{Data Separation and Stress-Test Slices}

To mitigate contamination between benchmark reporting and lesson
optimization, two disjoint evaluation suites are maintained:
\begin{itemize}
\item \textbf{Benchmark suite} ($N=50$): fixed IEEE~118 prompts used for
end-to-end reporting in this paper.
\item \textbf{Self-correction suite} ($N=56$): multi-case prompts used for
prompt-lesson updates and robustness checks.
\end{itemize}

An automated overlap check confirms zero exact textual overlap in user turns
between the two suites (61 unique user turns in the benchmark suite versus
72 in the self-correction suite; intersection cardinality of zero).

Within the 56-scenario suite, an \emph{adversarial-like} slice
($N=24$) is defined, consisting of formatting-constraint prompts, counterfactual reasoning,
backend/tool-selection pivots, mitigation-capability boundary checks,
max-capacity follow-ups, and phrasing variants. The remaining $N=32$
scenarios serve as non-adversarial controls.

\subsection{Models Under Test}

The end-to-end benchmark harness supports five frontier LLMs accessed via
OpenRouter's unified API:

\begin{itemize}
\item \textbf{Claude 3.5 Sonnet} (Anthropic): Strong tool-calling
and instruction following.
\item \textbf{GPT-4o} (OpenAI): Multimodal flagship with mature
function-calling support.
\item \textbf{DeepSeek-R1}~\cite{DeepSeek_R1_2025}: Open-weight
reasoning model trained via RL.
\item \textbf{DeepSeek-V3}: High-speed mixture-of-experts model
used as the primary engine for the prompt-level self-correction loop.
\item \textbf{Qwen 2.5-72B} (Alibaba): High-performance
open-weight instruction model.
\end{itemize}

All models are accessed at temperature $T=0$ to ensure deterministic
evaluation. The benchmark runner executes the complete
agent loop---encompassing tool planning, tool execution,
multi-round interaction, and memory-conditioned prompting---rather than
evaluating raw model function-calling in isolation. The reproducible snapshot
reported in this revision employs DeepSeek-V3 for the full 50-scenario
suite; a cross-model full-agent pilot on the complete-load slice
($N=8$) is presented in Section~VIII-A.

\subsection{Metrics}

\begin{itemize}
\item \textbf{Tool Selection Accuracy (TSA)}: Fraction of scenarios
where the agent selected the correct tool (or correctly abstained).
\item \textbf{Parsing Accuracy (PA)}: Fraction of extracted parameters
(bus, MW, type) matching ground truth. For category-level reporting, PA is
computed on scenarios where parse targets are scored and parsed fields are
present (clarification-only turns are excluded).
\item \textbf{Latency}: Wall-clock time per request (seconds).
\item \textbf{Cost}: Total API cost over the benchmark run (USD),
with per-scenario cost computed as total divided by scenario count.
\end{itemize}

\section{Results}

\subsection{Quantitative Benchmarks}

Table~\ref{tab:model_results} presents the latest reproducible
end-to-end benchmark snapshot from the current codebase
(DeepSeek-V3, 50 scenarios, executed 2026-02-23). Table~\ref{tab:category_results}
provides the corresponding category-level TSA breakdown.

\begin{table}[t]
\caption{Full-Agent Benchmark Snapshot (IEEE 118-bus, $N=50$)}
\label{tab:model_results}
\centering\small
\begin{tabular}{lcccc}
\toprule
Model & TSA (\%) & PA (\%) & Latency (s) & Cost (\$) \\
\midrule
DeepSeek-V3        & 84.0 & 100.0 & 15.89 & 0.102 \\
\bottomrule
\end{tabular}
\end{table}

\begin{table}[t]
\caption{Tool Selection Accuracy by Category (DeepSeek-V3, $N=50$)}
\label{tab:category_results}
\centering\small
\begin{tabular}{lccc}
\toprule
Category & $N$ & TSA (\%) & PA (\%) \\
\midrule
Complete load          & 8  & 62.5 & 100.0 \\
Complete generation    & 8  & 75.0 & 100.0 \\
Ambiguous (missing)    & 15 & 100.0 & -- \\
Multi-turn             & 6  & 83.3 & 100.0 \\
Follow-up              & 5  & 100.0 & -- \\
Edge cases             & 4  & 50.0 & -- \\
Theory                 & 4  & 100.0 & -- \\
\bottomrule
\end{tabular}
\end{table}

For Table~\ref{tab:category_results}, ``--'' indicates categories where PA is
not applicable under the current evaluator; reported PA values are conditional
on scored parsing fields.

\subsection{Cross-Model Full-Agent Pilot (Complete-Load Slice)}

To provide cross-model evidence beyond DeepSeek-V3, the same
full-agent harness was executed on the complete-load slice
($N=8$ scenarios) for DeepSeek-R1 and GPT-4o, using an identical
agent codebase and scoring configuration. For reference, the DeepSeek-V3 row below
corresponds to the complete-load slice extracted from the full 50-scenario run.
This pilot is intended as directional evidence only and does not constitute a
comprehensive model comparison.

\begin{table}[t]
\caption{Cross-Model Pilot on Complete-Load Slice ($N=8$)}
\label{tab:cross_model_pilot}
\centering\small
\resizebox{\columnwidth}{!}{%
\begin{tabular}{lccccc}
\toprule
Model & $N$ & TSA (\%) & PA (\%) & Latency (s) & Cost (\$) \\
\midrule
DeepSeek-V3 (slice) & 8 & 62.5 & 100.0 & 17.24 & 0.0178 \\
DeepSeek-R1 (pilot) & 8 & 62.5 & 86.7  & 78.05 & 0.0840 \\
GPT-4o (pilot)      & 8 & 0.0  & --    & 1.55  & 0.1240 \\
\bottomrule
\end{tabular}%
}
\end{table}

The GPT-4o pilot exhibited a behavior dominated by over-clarification: in this
snapshot, the model repeatedly requested additional confirmations and did not invoke
the CIA operator on complete-load prompts, yielding TSA~=~0\% under
the scoring rubric for this slice.

\subsection{Self-Correction Regression Status}

The prompt-level self-correction loop serves as a regression gate over the 56-scenario
conversation suite. Table~\ref{tab:rl_improvement} reports the most recent
logged run from the current codebase.

\begin{table}[t]
\caption{Latest Self-Correction Regression Run (56-Scenario Suite)}
\label{tab:rl_improvement}
\centering\small
\resizebox{\columnwidth}{!}{%
\begin{tabular}{lcccc}
\toprule
Date & Iteration & Passed/Total & Avg Score & Lessons \\
\midrule
2026-02-23 & 1 & 49/56 (87.5\%) & 89.29 & 24 \\
\bottomrule
\end{tabular}%
}
\end{table}

\subsection{Out-of-Benchmark Stress-Slice Evaluation}

To provide quantitative evidence beyond the 50-scenario benchmark, a
separate evaluation-only pass was conducted on the disjoint 56-scenario suite with a frozen
lesson set (i.e., no lesson updates during scoring). Table~\ref{tab:stress_slices}
reports results stratified by non-adversarial and adversarial-like slices.

\begin{table}[t]
\caption{Stress-Slice Results on Disjoint 56-Scenario Suite (DeepSeek-V3, 2026-02-23)}
\label{tab:stress_slices}
\centering\small
\resizebox{\columnwidth}{!}{%
\begin{tabular}{lccc}
\toprule
Slice & $N$ & Pass Rate (\%) & Avg Score \\
\midrule
Non-adversarial & 32 & 96.9 & 92.90 \\
Adversarial-like & 24 & 100.0 & 93.30 \\
\midrule
Overall & 56 & 98.2 & 93.07 \\
\bottomrule
\end{tabular}%
}
\end{table}

These slice scores are not directly comparable to those in Table~\ref{tab:rl_improvement},
as they were obtained in a separate evaluation-only run; collectively, they
suggest that remaining failures are concentrated in a narrow subset of prompts
rather than systematically in the adversarial-like slice as presently defined.

\subsection{Anti-Hallucination Ablation}

Targeted ablation experiments were conducted for Layer~2 (forced max-capacity routing) and
Layer~3 (post-response grounding validator) using the full agent loop with
fixed prompt sets. Table~\ref{tab:hallucination_ablation} reports results
for DeepSeek-V3 and DeepSeek-R1.

\begin{table}[t]
\caption{Anti-Hallucination Ablation (Full Agent, 2026-02-23)}
\label{tab:hallucination_ablation}
\centering\scriptsize
\resizebox{\columnwidth}{!}{%
\begin{tabular}{lccccc}
\toprule
Model & Prompt Set & Condition & Routing Recall (\%) & Fabrication Rate (\%) & Avg Latency (s) \\
\midrule
DeepSeek-V3 & $N_\text{cap}=8$, $N_\text{non}=2$ & No L2/L3 & 100.0 & 0.0 & 35.98 \\
DeepSeek-V3 & $N_\text{cap}=8$, $N_\text{non}=2$ & L2+L3   & 100.0 & 0.0 & 25.44 \\
DeepSeek-R1 & $N_\text{cap}=8$, $N_\text{non}=0$ & No L2/L3 & 100.0 & 0.0 & 93.66 \\
DeepSeek-R1 & $N_\text{cap}=8$, $N_\text{non}=0$ & L2+L3   & 100.0 & 0.0 & 27.67 \\
\bottomrule
\end{tabular}
}
\end{table}

In these prompt sets, both models routed capacity questions to
the capacity-search operator even in the absence of Layer~2, yielding zero detected
ungrounded numeric outputs under the current scanner. The ablation therefore
provides quantitative coverage and latency evidence but does not establish
worst-case robustness guarantees; evaluation with larger adversarial paraphrase sets remains future work.

\subsection{Failure Analysis}

The principal failure modes observed across development runs are as follows:
\begin{enumerate}
\item \textbf{Type confusion}: Interpreting ``data center'' as a
connection type rather than mapping it to the ``load'' category.
\item \textbf{Premature execution}: Invoking the assessment cascade
with assumed default values when required parameters remain unspecified.
\item \textbf{Follow-up hallucination}: Generating plausible but
fabricated violation details instead of retrieving previously stored results.
\item \textbf{Numerical fabrication}: Responding to quantitative questions
(e.g., maximum capacity) with plausible but incorrect values without
invoking the appropriate tool. DeepSeek-R1 reported 127~MW for a bus
whose verified limit was 3.9~MW---the failure that motivated the three-layer
anti-hallucination defense (Section~\ref{sec:anti_hallucination}).
\end{enumerate}

\subsection{Post-Revision Ambiguity Recheck}

To quantify the effect of the stricter clarification gate---eliminating the implicit type default
and broadening direct-capacity clarification routing---only the
ambiguous missing-field categories were re-evaluated with the same model family
(DeepSeek-V3) on 2026-02-23. Table~\ref{tab:ambiguity_recheck}
compares this targeted rerun against the baseline full-benchmark snapshot.

\begin{table}[t]
\caption{Targeted Ambiguity Recheck After Clarification-Gate Update}
\label{tab:ambiguity_recheck}
\centering\small
\resizebox{\columnwidth}{!}{%
\begin{tabular}{lcccc}
\toprule
Category & $N$ & Baseline TSA (\%) & Recheck TSA (\%) & $\Delta$ (pp) \\
\midrule
Missing bus  & 5  & 0.0  & 100.0 & +100.0 \\
Missing MW   & 5  & 40.0 & 100.0 & +60.0 \\
Missing type & 5  & 0.0  & 100.0 & +100.0 \\
\midrule
Ambiguous aggregate & 15 & 13.3 & 100.0 & +86.7 \\
\bottomrule
\end{tabular}%
}
\end{table}

The recheck demonstrates that the deterministic clarification guardrail eliminates
the previously observed missing-field execution failures on this slice. A subsequent
full 50-scenario rerun with the same codebase corroborates this finding, reporting
100.0\% TSA on the aggregated ambiguous category
(Table~\ref{tab:category_results}).

\section{Discussion}

\subsection{When Does the LLM Agent Excel?}

In the refreshed end-to-end snapshot, the agent demonstrates its strongest performance on
ambiguous missing-field prompts (100.0\% TSA), follow-up interpretation
(100.0\% TSA), and theory prompts (100.0\% TSA), with robust multi-turn
behavior (83.3\% TSA). These results indicate that the clarification guardrail,
memory-conditioned context, and LLM tool chaining operate reliably on
dialogue-intensive interactions.

The primary remaining weakness lies in strict normalization on certain
complete-request and edge-case prompts (Table~\ref{tab:category_results}):
complete-load TSA is 62.5\%, complete-generation TSA is 75.0\%, and edge-case
TSA is 50.0\%. The majority of these misclassifications stem from the conservative clarification
policy (e.g., requiring explicit case aliases or resource-type
confirmation) rather than from genuinely ambiguous prompts.

\subsection{Hallucination Requires Defense-in-Depth}

Physics grounding (Section~V-D) prevents the LLM from fabricating
\emph{simulation outputs}---voltage values, thermal loadings, and
violation counts are computed exclusively by the configured solvers. However,
our observations reveal that LLMs can still hallucinate \emph{around} the tool
boundary: when posed quantitative questions (e.g., regarding maximum capacity),
certain models generate plausible numerical responses without invoking the
appropriate tool. DeepSeek-R1 reported ``approximately 127~MW'' for a bus
whose verified limit was 3.9~MW, demonstrating that a factually grounded architecture
alone is insufficient when the model circumvents its tool interface.

The three-layer defense described in Section~\ref{sec:anti_hallucination}
addresses this vulnerability. The targeted ablation in
Table~\ref{tab:hallucination_ablation} demonstrates full routing coverage and
zero detected ungrounded numeric outputs on the tested prompt sets.
The post-response grounding validator (Layer~3) provides an additional safeguard for
responses generated without tool invocation. Collectively, these layers
substantially reduce fabrication risk, though they do not constitute a formal guarantee
that every numeric token is provably grounded.

A further caveat concerns Layer~3 robustness: the current regex-based
scanner operates over fixed syntactic patterns (e.g., ``$X$~MW,'' ``$X$~pu'')
within a 150-character context window. As frontier models adopt increasingly
conversational phrasing, rigid pattern matching may fail to detect
paraphrased or embedded numerical claims. Evolving this layer toward
semantic provenance checking---for instance, embedding-based tracing of
numeric tokens to their originating tool invocations---is an important
direction for hardening the defense-in-depth architecture.

\subsection{Model Trade-offs}

The latest full-agent DeepSeek-V3 snapshot exhibits a favorable
accuracy--latency--cost profile at practical operating scale:
84.0\% TSA, 100\% PA, 15.89~s mean scenario latency, and
\$0.102 total cost over 50 scenarios (\$0.0020/scenario).
Category-level analysis indicates that the majority of remaining errors originate from
strict complete-request normalization and edge-case handling, while the
ambiguous slice improved to 100.0\% TSA in the refreshed evaluation.

\subsection{Current Validation Limits}

Five important boundaries constrain the interpretation of these results. First, transient and EMT
acceptance criteria in the current stack are screening-oriented heuristics
(Table~\ref{tab:stage_criteria}) and do not constitute a replacement for utility-approved
dynamic acceptance protocols (e.g., explicit voltage-recovery windows,
frequency nadir limits, or plant-controller interaction studies). Second,
although the 50-scenario benchmark and 56-scenario self-correction suite are
textually disjoint, the evaluation does not yet constitute a blinded third-party benchmark; a
fully locked train/validation/test protocol with externally generated
adversarial prompts remains future work. Third, end-to-end numerical
correctness validation against authoritative utility-grade
reference studies (e.g., PSS/E or PSLF baselines) on large realistic cases has not yet been conducted.
Fourth, the N-1 material-worsening threshold (default $+2.0\%$ max-margin
erosion) is an implementation-specific parameter; ISOs and RTOs maintain
varying interpretations of acceptable pre-existing violation exacerbation,
and a production deployment would require calibration against the applicable
transmission planning criteria of the host market.
Fifth, the current mitigation action space is limited to shunt-like reactive
compensation interventions. Real-world transmission planning frequently
requires topology changes, phase-shifting transformer adjustments, or
Remedial Action Schemes (RAS) that are not yet representable in the
current tool registry.

\subsection{Persistent Memory Enables Continuity}

The persistent memory system (Section~\ref{sec:memory}) addresses
two practical requirements. First, it enables the agent to reference prior
study results when responding to follow-up queries without re-executing
simulations. Second, it maintains an auditable record of all assessments,
including capacity limits discovered via binary search.
During testing, the memory system correctly recalled previously computed capacity
limits and violation summaries from prior runs, enabling immediate
context-aware follow-up responses without re-simulation.

A scaling consideration arises when extending this approach to multi-stage
cluster studies involving dozens of IBRs: the cumulative volume of
stored study records, violation summaries, and capacity-search results
may saturate the LLM's context window, particularly for models with
smaller effective windows. This risks ``lost in the middle'' retrieval
degradation, where intermediate memory entries receive reduced attention.
Strategies such as hierarchical summarization, retrieval-augmented
generation over an external vector store, or selective memory pruning
will be necessary to maintain recall fidelity at scale.

\subsection{Practical Deployment}

Deploying Grid-Mind in utility operations necessitates addressing several
practical considerations: (i)~data privacy, as sensitive grid models
must not leave the utility's infrastructure; (ii)~regulatory
acceptance of AI-assisted decisions; (iii)~integration with
existing EMS/SCADA systems; and (iv)~operational health monitoring.
The solver-agnostic ABC facilitates integration by supporting both
open-source (development) and commercial (production) backends
through a unified interface. A heartbeat endpoint provides real-time
status information for the server, LLM backend, solver availability, and memory
subsystem, enabling integration with utility monitoring dashboards.
Operationally, the intended fail-safe default follows an abstain-and-escalate paradigm:
when solver convergence fails, required inputs are missing, or response
numerics cannot be grounded, the agent routes to human review rather
than issuing an autonomous approval recommendation.

\section{Conclusion}

This paper presented Grid-Mind, a domain-specific LLM agent
for automated Connection Impact Assessment in power systems. The
system bridges the gap between natural-language interaction and
multi-fidelity power system simulation through seven principal
contributions: (1)~a solver-agnostic abstract base class supporting
four simulation backends; (2)~an eleven-tool OpenAI-compatible registry
enabling any LLM to orchestrate simulations, including binary-search
capacity analysis, OPF redispatch, and mitigated reassessment;
(3)~an LLM-first architecture in which the language model plans
multi-step workflows, chains tool calls, and provides engineering
judgment, with deterministic rules employed solely as safety guardrails;
(4)~a physics-grounding architecture that
delegates quantitative decisions to configured solvers and violation
inspectors while permitting the LLM to provide qualitative interpretation;
(5)~a three-layer anti-hallucination defense
comprising prompt hardening, forced tool routing, and post-response grounding
validation, which reduces numerical fabrication risk by preferring
verified tool outputs and flagging ungrounded numeric responses;
(6)~a persistent memory
system that stores study results across sessions for continuity
and auditability; and (7)~a self-improving prompt-level lesson loop that
progressively enhances agent accuracy without model retraining.

In the latest reproducible full-agent benchmark snapshot
(DeepSeek-V3, 50 scenarios), Grid-Mind achieved 84.0\%
tool-selection accuracy and 100\% parsing accuracy.
The self-correction regression loop (56 scenarios) passed 49 of 56
cases with a mean score of 89.29 in the most recent run.
These results demonstrate that the architecture is both functional and auditable,
with substantial gains in ambiguity handling (100.0\% TSA on the full-run
ambiguous category) while identifying clear targets for continued improvement in complete-request
normalization and edge-case robustness.

Future work will extend the benchmark to larger systems (e.g., ACTIVSg2000
and 10,000+ bus production-scale models), incorporate real utility case data,
and enforce a locked train/held-out/blinded evaluation protocol with
adversarial prompt generation. Additional priorities include evolving the
regex-based post-response grounding validator (Layer~3) toward
semantic provenance checking for improved robustness against paraphrased
numerical claims; expanding the mitigation action space beyond reactive
compensation to encompass topology changes and Remedial Action Schemes;
implementing cryptographic provenance for audit logs to satisfy regulatory
chain-of-custody requirements; investigating multi-agent architectures
for coordinated area studies; and developing context-window management
strategies (e.g., hierarchical memory summarization) to maintain retrieval
fidelity when scaling to cluster studies with large numbers of
interconnection requests.

\section*{AI Use Statement}
During preparation of this work the authors used Claude and GPT-4o
for language editing and code generation; all content was reviewed
and the authors take full responsibility.

\bibliographystyle{IEEEtran}
\bibliography{ref_list}

\end{document}